\documentclass[seceq]{ptptex}

\usepackage{graphicx}
\usepackage{wrapft}

\title{%        %You can use \\ for explicit line-break.
Quantum computation with the Jaynes-Cummings model%
}

\author{%       %Use \scshape for the family name.
Hiroo \textsc{Azuma}$^{}$\footnote{Present address:
Advanced Algorithm \& Systems Co., Ltd.,
7F Ebisu-IS Building, 1-13-6 Ebisu, Shibuya-ku, Tokyo 150-0013, Japan.
\\
E-mail: hiroo.azuma@m3.dion.ne.jp} %
}

\inst{%     %Affiliation, neglected when [addenda] or [errata].
Institute of Computational Fluid Dynamics,
1-16-5 Haramachi, Meguro-ku, Tokyo 152-0011, Japan
}

\abst{%         %This abstract is neglected when [addenda] or [errata].
In this paper, we propose a method for building a two-qubit gate
with the Jaynes-Cummings model (JCM).
In our scheme, we construct a qubit from a pair of optical paths
where a photon is running.
Generating Knill, Laflamme and Milburn's nonlinear sign-shift gate
by the JCM,
we construct the conditional sign-flip gate,
which works with small error probability in principle.
We also discuss two experimental setups for realizing our scheme.
In the first experimental setup, we make use of coherent lights to examine
whether or not our scheme works.
In the second experimental setup, an optical loop circuit
made out of the polarizing beam splitter
and the Pockels cell takes an important role in the cavity.
}

\PTPindex{061}  %Input the subject index(es) of your paper, 
\begin{document}

\maketitle

\section{\label{introduction}Introduction}
Since Shor's quantum algorithm for factoring large integers more efficiently
than classical algorithms
and
Grover's efficient amplitude amplification process
for quantum states appeared,\cite{Shor1997,Grover1997}
experimental realization of quantum computation has been attracting
many physicists' attention.
Quantum computation is a sequence of the following operations.
First,
we prepare a superposition of quantum states as an input with quantum bits
(two-state quantum systems, namely qubits).
Second,
we apply unitary transformations to these qubits
by quantum logic gates.
Finally,
to obtain an output,
we observe the qubits with appropriate measurement basis vectors.
Hence,
to implement a quantum computer,
we have to build the qubits and the quantum logic gates.

In general,
we can construct an arbitrary one-qubit gate
which applies a unitary transformation
to a single qubit at ease,
no matter which physical system we choose as the qubit.
In contrast,
implementation of a two-qubit gate is thought to be very difficult
because it has to generate nonlocal quantum correlation (entanglement)
between two local qubits.
(Entanglement is regarded as a resource of quantum information processing.)
Moreover, it is shown
that we can construct any unitary transformation applied to an arbitrary number of qubits
from the one-qubit gates and a certain two-qubit gate,
such as the controlled-NOT gate
and the conditional sign-flip gate.\cite{Barenco1995} \ 
Because of these reasons,
implementation of the two-qubit gate is
one of the most important points of quantum computation.

Here we propose how to construct the conditional sign-flip gate
with the Jaynes-Cummings model (JCM),
which is a quantum mechanical model describing the interaction
between a single two-level atom and a single electromagnetic field mode.
We also give discussions about experimental setups for realizing our scheme.

Properties of the JCM have been studied theoretically from
1960s and they were confirmed by the experiment
in 1980s,\cite{Jaynes1963,Shore1993,Louisell1973,Walls1994,Schleich2001,Rempe1987}
so that the JCM is familiar to and well-studied by the researchers
in the field of atomic and optical physics.
Furthermore,
these days,
the JCM is used for describing the evolution of entanglement
between the atom and the photons
by researchers of quantum information science.\cite{Bose2001,Scheel2003} \ 
[Azuma investigates the thermal JCM and discusses the lower bound
of entanglement between the two-level atom and thermal photons.\cite{Azuma2008-1}]
Recently, so-called sudden death effect
(disappearance of entanglement of two isolated Jaynes-Cummings atoms
in a finite time)
has been studied eagerly.\cite{Yu2004,Yonac2006,Yu2006} \ 
This phenomenon is experimentally demonstrated.\cite{Almeida2007} \ 
Thus, we can expect that our proposal grows to be one of the various applications
of the JCM.
Moreover, we can expect that our proposal becomes one of the important candidates
for the method of implementing quantum logic,
for example,
cold trapped ions interacting with laser beams,\cite{Cirac1995,Monroe1995}
polarized photons in the cavity quantum electrodynamics
system,\cite{Turchette1995}
and so on.

As mentioned above,
to construct a quantum computer,
we have to prepare qubits and quantum logic gates.
In our method, we regard a pair of optical paths
where a photon is running as a qubit.
[This construction of a qubit is called
the dual-rail qubit representation.\cite{Chuang1995}]
We can apply any one-qubit transformation to this dual-rail qubit
with beam splitters and phase shifters.

Knill, Laflamme and Milburn (KLM) show a unique method
of applying the conditional sign-flip gate to two dual-rail qubits
using beam splitters and the nonlinear sign-shift (NS) gate,\cite{Knill2001}
which causes the following transformation
to the number states of photons:
\begin{equation}
\alpha|0\rangle_{\mbox{\scriptsize P}}
+\beta|1\rangle_{\mbox{\scriptsize P}}
+\gamma|2\rangle_{\mbox{\scriptsize P}}
\rightarrow
\alpha|0\rangle_{\mbox{\scriptsize P}}
+\beta|1\rangle_{\mbox{\scriptsize P}}
-\gamma|2\rangle_{\mbox{\scriptsize P}}.
\label{NS-gate-transformation}
\end{equation}
(The index P stands for the photons.)

In this paper, we show the method of implementing the NS gate with the JCM.
In KLM's proposal, the NS gate is constructed only with passive linear optics,
and it works
as a nondeterministic gate conditioned on the detection of an auxiliary photon.
(It works with probability $1/4$.)
Someone may disagree with our proposal
because we are going to introduce a nonlinear device into KLM's scheme.
However, in our method, the NS gate works with small error probability in principle.
Thus, the author thinks that our method is a practical solution
for the simplification of the whole system of the NS gate.

Gilchrist {\it et al}. try to build the NS gate by trapped atoms
in an optical cavity.\cite{Gilchrist2003} \ 
Marchiolli {\it et al}. investigate the JCM
with an external field and give qualitative analyses
about the entanglement between the atom and the cavity field.\cite{Marchiolli2003,Marchiolli2006} \ 
These studies seem to relate to our scheme.
Azuma proposes a method of constructing the NS gate with one-dimensional
Kerr-nonlinear photonic crystals.\cite{Azuma2008-2} \ 
That work aims to implement the NS gate with simple structures of matters.
Reference~\citen{Azuma2008-2} and our scheme explained in this paper share
an idea of building the NS gate with a practical method.

In this paper,
we concentrate our attention on constructing a single two-qubit gate
that is realized by nonlinear interaction and works with high fidelity.
Recently, tackling the subject in a different way,
some researchers in the field of optical quantum computation
have been making new developments towards quantum gates,
which work with some fidelity (not high-performance)
and are practical for large scale computation.
Stephens {\it et al}. discuss a large scale deterministic optical quantum computer,
which utilizes atom-cavity Q-switches and two-dimensional
cluster states.\cite{Stephens2008}

In the latter half of this paper,
we discuss two experimental setups
for realizing our schemes.
In the second experimental setup,
we try to construct a high-Q cavity
where the atom and the cavity-mode prepared
as an input for the quantum logic interact with each other efficiently.
In this setup,
we build an optical loop circuit
made out of the polarizing beam splitter and the Pockels cell.
Because construction of the high quality Q-switch is
one of important topics for the optical quantum computation,
many ideas for high-Q cavity are proposed.
Birnbaum {\it et al}. experimentally demonstrate photon blockade
in an optical cavity with a single trapped atom.\cite{Birnbaum2005} \ 
In their experiment,
the first photon in the cavity forbids the transmission of the second photon,
so that the atom in the cavity interacts with photons one-by-one.

This paper is organized as follows.
In section~\ref{dual-rail-qubits-KLM-scheme},
we explain the dual-rail qubit representation and KLM's scheme.
[This section is a short review of references~\citen{Chuang1995} and \citen{Knill2001}.]
In section~\ref{construction-NS-gate-JCM},
we explain how to build the NS gate by the JCM.
In section~\ref{an-experimental-setup1},
we discuss an experimental setup in which we make use of coherent lights
to examine whether or not our scheme works.
In section~\ref{an-experimental-setup2},
we discuss another experimental setup in which an optical loop circuit
made out of a polarizing beam splitter and the Pockels cell
takes an important role in the cavity.
In section~\ref{discussions}, we give brief discussions.

\section{\label{dual-rail-qubits-KLM-scheme}Dual-rail qubits and KLM's scheme}
In this section, we explain the dual-rail qubit representation and KLM's scheme.
[This section is a short review of references~\citen{Chuang1995} and \citen{Knill2001}.
The facts described in this section are utilized in reference~\citen{Azuma2008-2},
as well.]
First, we build a qubit by the dual-rail qubit representation.
First of all,
we prepare a pair of optical paths, $x1$ and $x2$.
Each optical path can take a superposition
of the number states
$|n\rangle_{\mbox{\scriptsize P}}$ for $n=0,1,2,...$,
where $n$ is the number of photons on the path.
Then,
$|0\rangle_{x1}\otimes|1\rangle_{x2}$ is a state
where paths $x1$ and $x2$ have zero and one photons, respectively,
and we regard it as a logical ket vector $|\bar{0}\rangle_{x}$.
We regard $|1\rangle_{x1}\otimes|0\rangle_{x2}$
as a logical ket vector $|\bar{1}\rangle_{x}$, similarly.
And we describe an arbitrary state of a qubit
as $|\phi\rangle_{x}
=\alpha|\bar{0}\rangle_{x}+\beta|\bar{1}\rangle_{x}$.

\begin{figure}
\centerline{\includegraphics[scale=1.0]{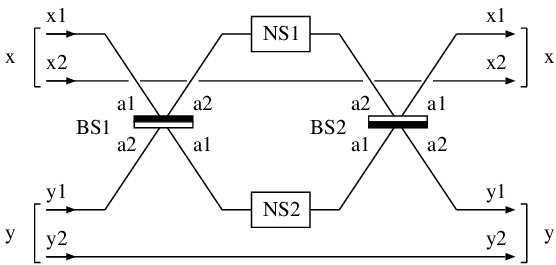}}
\caption{Implementation of the conditional sign-flip gate
with the NS gates
defined in equation~(\ref{NS-gate-transformation}).
Qubits $x$ and $y$ consist of a pair of paths $x1$ and $x2$
and a pair of paths $y1$ and $y2$, respectively.
Symbols BS1 and BS2 represent beam splitters (half-silvered mirrors).
Symbols NS1 and NS2 represent the NS gates.
Photons travel from left to right in this network.}
\label{Figure01}
\end{figure}

Next, we construct the conditional sign-flip gate with the NS gates
defined in equation~(\ref{NS-gate-transformation}).
An optical network drawn in figure~\ref{Figure01} works
as the conditional sign-flip gate,
whose operation is given by
$|\bar{j}\rangle_{x}\otimes|\bar{k}\rangle_{y}\rightarrow
(-1)^{jk}|\bar{j}\rangle_{x}\otimes|\bar{k}\rangle_{y}$
for $j,k\in\{0,1\}$.
Let us confirm the function of this network.
In figure~\ref{Figure01}, symbols BS1 and BS2 represent beam splitters
(half-silvered mirrors),
which transform the incident number states of paths $a1$ and $a2$ as follows:
\begin{eqnarray}
&&
|n\rangle_{a1}|m\rangle_{a2}
=
\frac{1}{\sqrt{n!m!}}(a_{1}^{\dagger})^{n}(a_{2}^{\dagger})^{m}
|0\rangle_{a1}|0\rangle_{a2} \nonumber \\
&\rightarrow&
\frac{1}{\sqrt{n!m!}}
[\frac{1}{\sqrt{2}}(a_{1}^{\dagger}+a_{2}^{\dagger})]^{n}
[\frac{1}{\sqrt{2}}(a_{1}^{\dagger}-a_{2}^{\dagger})]^{m}
|0\rangle_{a1}|0\rangle_{a2} \nonumber \\
&&\quad\quad\mbox{for $n,m\in\{0,1,2,...\}$},
\label{function-beamsplitter}
\end{eqnarray}
where $a_{1}^{\dagger}$ and $a_{2}^{\dagger}$ are
creation operators of photons on the paths $a1$ and $a2$, respectively.
We give attention to the fact that BS2 applies an inverse transformation
of BS1.
Symbols NS1 and NS2 represent the NS gates.

Putting a superposition of $|\bar{0}\rangle_{x}|\bar{0}\rangle_{y}$,
$|\bar{0}\rangle_{x}|\bar{1}\rangle_{y}$
and
$|\bar{1}\rangle_{x}|\bar{0}\rangle_{y}$
into the left side of the network shown in figure~\ref{Figure01},
the network leaves it untouched and returns it
as an output from the right side of the network.
In contrast, if we put a state
$|\bar{1}\rangle_{x}|\bar{1}\rangle_{y}
=|1\rangle_{x1}|0\rangle_{x2}
|1\rangle_{y1}|0\rangle_{y2}$
into the network,
the following transformation is applied to the paths $x1$ and $y1$:
\begin{eqnarray}
|1\rangle_{x1}|1\rangle_{y1}
&\stackrel{\mbox{\scriptsize BS1}}{\longrightarrow}&
\frac{1}{\sqrt{2}}
(|2\rangle_{a1}|0\rangle_{a2}
-|0\rangle_{a1}|2\rangle_{a2}) \nonumber \\
&\stackrel{\mbox{\scriptsize NS1,NS2}}{\longrightarrow}&
-\frac{1}{\sqrt{2}}
(|2\rangle_{a1}|0\rangle_{a2}
-|0\rangle_{a1}|2\rangle_{a2}) \nonumber \\
&\stackrel{\mbox{\scriptsize BS2}}{\longrightarrow}&
-|1\rangle_{x1}|1\rangle_{y1}.
\label{the-function-of-network}
\end{eqnarray}
Thus, we obtain
$-|\bar{1}\rangle_{x}|\bar{1}\rangle_{y}
=
-|1\rangle_{x1}|0\rangle_{x2}
|1\rangle_{y1}|0\rangle_{y2}$
as an output for the input state
$|\bar{1}\rangle_{x}|\bar{1}\rangle_{y}$.
Hence, the network shown in figure~\ref{Figure01}
realizes the conditional sign-flip gate.

\section{\label{construction-NS-gate-JCM}Construction of the NS gate with the JCM}
The JCM is originally designed for describing a spontaneous emission of the atom.
Its Hamiltonian is given by
$H=\hbar(C_{1}+C_{2})$,
$C_{1}=\omega[(1/2)\sigma_{z}+a^{\dagger}a]$,
$C_{2}=\kappa(\sigma_{+}a+\sigma_{-}a^{\dagger})$,
where
$\sigma_{\pm}=(1/2)(\sigma_{x}\pm i\sigma_{y})$
and
$[a,a^{\dagger}]=1$.
The Pauli matrices ($\sigma_{i}$, $i=x,y,z$) are operators of the atom, and
$a$ and $a^{\dagger}$ are annihilation and creation operators
of the electromagnetic field, respectively.
Here, we assume that $\kappa$ is a real constant
and the field is resonant with the atom.
(The photons' frequency is equal to the energy gap of the two-level atom.)

Because $[C_{1},C_{2}]=0$ and $C_{1}$ can be diagonalized at ease,
we take the following interaction picture.
We write a state vector of the whole system in the Schr{\"o}dinger picture
as $|\psi_{\mbox{\scriptsize S}}(t)\rangle$.
A state vector in the interaction picture is defined by
$|\psi_{\mbox{\scriptsize I}}(t)\rangle=\exp(iC_{1}t)|\psi_{\mbox{\scriptsize S}}(t)\rangle$.
[We assume $|\psi_{\mbox{\scriptsize I}}(0)\rangle=|\psi_{\mbox{\scriptsize S}}(0)\rangle$.]
The time evolution of $|\psi_{\mbox{\scriptsize I}}(t)\rangle$ is given by
$|\psi_{\mbox{\scriptsize I}}(t)\rangle=U(t)|\psi_{\mbox{\scriptsize I}}(0)\rangle$,
where $U(t)=\exp(-iC_{2}t)$.

We define the basis vectors for the states of the atom and the photons
as follows.
The ground and excited states of the atom are given by two-component vectors,
\begin{equation}
|g\rangle_{\mbox{\scriptsize A}}
=
\left(
\begin{array}{c}
0 \\
1
\end{array}
\right),
\quad\quad
|e\rangle_{\mbox{\scriptsize A}}
=
\left(
\begin{array}{c}
1 \\
0
\end{array}
\right),
\label{atom-basis-vectors}
\end{equation}
where we assume that $|g\rangle_{\mbox{\scriptsize A}}$ and $|e\rangle_{\mbox{\scriptsize A}}$ are
eigenvectors of $\sigma_{z}$
with eigenvalues $-1$ and $1$, respectively.
(The index A stands for the atom.)
The number states of the photons are described as
$|n\rangle_{\mbox{\scriptsize P}}$ ($n=0,1,2,...$).

Describing the atom's Pauli operators by $2\times 2$ matrices,
we can write down $U(t)$ as follows:
\begin{equation}
U(t)=\exp[-it
\left(
\begin{array}{cc}
0 & \kappa a \\
\kappa a^{\dagger} & 0
\end{array}
\right)
]
=
\left(
\begin{array}{cc}
u_{00} & u_{01} \\
u_{10} & u_{11}
\end{array}
\right),
\label{unitary-evolution-1}
\end{equation}
where
\begin{eqnarray}
u_{00}&=&\cos(|\kappa|t\sqrt{a^{\dagger}a+1}), \nonumber \\
u_{01}&=&-i\kappa a
\frac{\sin(|\kappa|t\sqrt{a^{\dagger}a})}{|\kappa|\sqrt{a^{\dagger}a}},
\nonumber \\
u_{10}&=&-i\kappa a^{\dagger}
\frac{\sin(|\kappa|t\sqrt{a^{\dagger}a+1})}{|\kappa|\sqrt{a^{\dagger}a+1}},
\nonumber \\
u_{11}&=&\cos(|\kappa|t\sqrt{a^{\dagger}a}).
\label{unitary-evolution-2} 
\end{eqnarray}
The time evolution of the three initial states,
$|\psi_{\mbox{\scriptsize I}}(0)\rangle
=|g\rangle_{\mbox{\scriptsize A}}|0\rangle_{\mbox{\scriptsize P}}$,
$|g\rangle_{\mbox{\scriptsize A}}|1\rangle_{\mbox{\scriptsize P}}$
and
$|g\rangle_{\mbox{\scriptsize A}}|2\rangle_{\mbox{\scriptsize P}}$
are given by
\begin{eqnarray}
U(t)|g\rangle_{\mbox{\scriptsize A}}|0\rangle_{\mbox{\scriptsize P}}
&=&
U(t)
\left(
\begin{array}{c}
0 \\
|0\rangle_{\mbox{\scriptsize P}}
\end{array}
\right)
=
\left(
\begin{array}{c}
0 \\
|0\rangle_{\mbox{\scriptsize P}}
\end{array}
\right), \nonumber \\
U(t)|g\rangle_{\mbox{\scriptsize A}}|1\rangle_{\mbox{\scriptsize P}}
&=&
U(t)
\left(
\begin{array}{c}
0 \\
|1\rangle_{\mbox{\scriptsize P}}
\end{array}
\right) \nonumber \\
&=&
\left(
\begin{array}{c}
-i (\kappa/|\kappa|)\sin(|\kappa|t)|0\rangle_{\mbox{\scriptsize P}} \\
\cos(|\kappa|t)|1\rangle_{\mbox{\scriptsize P}}
\end{array}
\right), \nonumber \\
U(t)|g\rangle_{\mbox{\scriptsize A}}|2\rangle_{\mbox{\scriptsize P}}
&=&
U(t)
\left(
\begin{array}{c}
0 \\
|2\rangle_{\mbox{\scriptsize P}}
\end{array}
\right) \nonumber \\
&=&
\left(
\begin{array}{c}
-i (\kappa/|\kappa|)\sin(\sqrt{2}|\kappa|t)|1\rangle_{\mbox{\scriptsize P}} \\
\cos(\sqrt{2}|\kappa|t)|2\rangle_{\mbox{\scriptsize P}}
\end{array}
\right).
\label{unitary-evolution-3}
\end{eqnarray}

We want to obtain the NS gate
given by equation~(\ref{NS-gate-transformation}),
which flips only the sign of the coefficient of
$|2\rangle_{\mbox{\scriptsize P}}$.
Thus, we let $t=(2m+1)\pi/(\sqrt{2}|\kappa|)$ for $m=0,1,2,...$,
and we obtain the following time evolution:
$|g\rangle_{\mbox{\scriptsize A}}|0\rangle_{\mbox{\scriptsize P}}
\rightarrow
|g\rangle_{\mbox{\scriptsize A}}|0\rangle_{\mbox{\scriptsize P}}$,
$|g\rangle_{\mbox{\scriptsize A}}|1\rangle_{\mbox{\scriptsize P}}
\rightarrow
c(m)|e\rangle_{\mbox{\scriptsize A}}|0\rangle_{\mbox{\scriptsize P}}
+d(m)|g\rangle_{\mbox{\scriptsize A}}|1\rangle_{\mbox{\scriptsize P}}$,
$|g\rangle_{\mbox{\scriptsize A}}|2\rangle_{\mbox{\scriptsize P}}
\rightarrow
-|g\rangle_{\mbox{\scriptsize A}}|2\rangle_{\mbox{\scriptsize P}}$,
where
\\
$c(m)=-i(\kappa/|\kappa|)\sin[(2m+1)\pi/\sqrt{2}]$
and $d(m)=\cos[(2m+1)\pi/\sqrt{2}]$.

\begin{wraptable}{l}{\halftext}
\caption{The variation of the error probability $|c(m)|^{2}$
and the coefficient of $|g\rangle_{\mbox{\scriptsize A}}|1\rangle_{\mbox{\scriptsize P}}$
of the evolved state $d(m)$.}
\label{Table1}
\begin{center}
\begin{tabular}{lll} \hline \hline
$m$ & $|c(m)|^{2}$ & $d(m)$ \\
\hline
$0$ & $0.633$ & $-0.606$ \\
$1$ & $0.138$ & $0.928$ \\
$2$ & $0.988$ & $0.111$ \\
$3$ & $0.0247$ & $-0.988$ \\
$4$ & $0.828$ & $0.414$ \\
\hline
\end{tabular}
\end{center}
\end{wraptable}

In table~\ref{Table1}, we show values of $|c(m)|^{2}$ and $d(m)$
for $m=0,1,2,3,4$.
We pay attention to the cases of $m=1$ and $m=3$.
When we let $m=1$, $|c(1)|^{2}$ takes a small value
and $d(1)$ is nearly equal to unity.
This implies that if we let $t=3\pi/(\sqrt{2}|\kappa|)$,
we obtain the operation of the NS gate shown
in equation~(\ref{NS-gate-transformation})
with the error probability upper bound $0.138$.
When we let $m=3$, $|c(3)|^{2}$ is nearly equal to zero
and $d(m)$ is nearly equal to $-1$.
This implies that if we let $t=7\pi/(\sqrt{2}|\kappa|)$,
we obtain the transformation,
$\alpha|0\rangle_{\mbox{\scriptsize P}}+\beta|1\rangle_{\mbox{\scriptsize P}}
+\gamma|2\rangle_{\mbox{\scriptsize P}}
\rightarrow
\alpha|0\rangle_{\mbox{\scriptsize P}}-\beta|1\rangle_{\mbox{\scriptsize P}}
-\gamma|2\rangle_{\mbox{\scriptsize P}}$
with the error probability upper bound $0.0247$.
To flip the sign of the coefficient of $|1\rangle_{\mbox{\scriptsize P}}$
after this transformation,
we apply a phase shifter to the photons.
This device causes the operation,
$|n\rangle_{\mbox{\scriptsize P}}\rightarrow(-1)^{n}|n\rangle_{\mbox{\scriptsize P}}$,
so that we obtain the transformation,
$\alpha|0\rangle_{\mbox{\scriptsize P}}-\beta|1\rangle_{\mbox{\scriptsize P}}
-\gamma|2\rangle_{\mbox{\scriptsize P}}
\rightarrow
\alpha|0\rangle_{\mbox{\scriptsize P}}+\beta|1\rangle_{\mbox{\scriptsize P}}
-\gamma|2\rangle_{\mbox{\scriptsize P}}$,
and we finally obtain the transformation of the NS gate shown
in equation~(\ref{NS-gate-transformation})
with the error probability upper bound $0.0247$.

From the above method,
we obtain the NS gate which works
with the intrinsic error probability $0.138$ or $0.0247$.
Thus,
building a quantum circuit with our NS gate
for performing practical and robust quantum computation,
we have to utilize the quantum error correcting codes.
The quantum error correction and the fault-tolerant quantum computing
have been discussed eagerly
and
they are established theoretically.\cite{Shor1995,Steane1996,Calderbank1996,Shor1996,DiVincenzo1996}

For example,
if we use the Calderbank-Shor-Steane code,
which maps one qubit into seven qubits
and corrects an arbitrary one-qubit error in a block of qubits,
we can obtain the threshold of the error probability $1/7(\simeq 0.143)$ around
for each qubit.
Thus,
we can overcome the intrinsic error probability of our NS gate in principle.

\section{\label{an-experimental-setup1}An experimental setup for examining the function
of the NS gate using coherent lights}
In our scheme explained in section~\ref{construction-NS-gate-JCM},
we have to put the photons' initial state,
$\alpha|0\rangle_{\mbox{\scriptsize P}}+\beta|1\rangle_{\mbox{\scriptsize P}}
+\gamma|2\rangle_{\mbox{\scriptsize P}}$,
into the cavity and let it develop into the cavity mode.
Then, we have to let the cavity mode interact with the two-level atom as the JCM,
and finally extract the evolved state of photons from the cavity.
In general, it is difficult to perform these sequential procedures
practically in the laboratory.

\begin{figure}
\centerline{\includegraphics[scale=1.0]{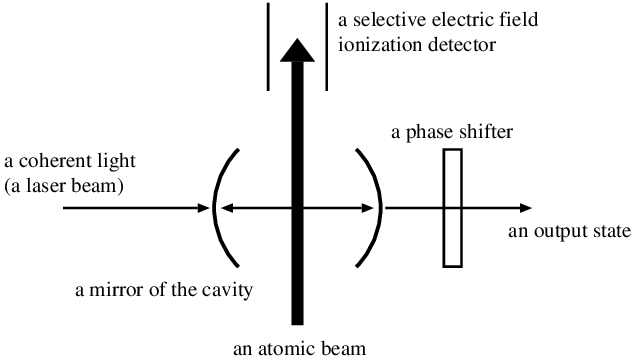}}
\caption{An outline of the experimental setup which examines whether or not
the NS gate realized by our scheme works.
A coherent light (a laser beam) is fed to the cavity from its left side
and reflected by mirrors of the cavity many times,
so that it becomes a cavity field.
The two-level atom passes the cavity as a slow beam and it causes the Jaynes-Cummings interaction
with the cavity field.
Then, the time-evolved state of the cavity field runs away from the cavity to its right side.
[Strictly speaking, the time-evolved state of the cavity field can fly away from the cavity
to its either side (the right side or the left side).
However, to let the discussion be simple,
we assume that the time-evolved light goes outside of the cavity from its right side in this figure.]
If the time of flight of the atom through the cavity is given by
$T=3\pi/(\sqrt{2}|\kappa|)$,
we need not put a phase shifter behind the cavity.
On the other hand, if $T=7\pi/(\sqrt{2}|\kappa|)$, we have to put it there.
A selective electric field ionization detector distinguishes the atom's ground state
$|g\rangle_{\mbox{\scriptsize A}}$
from its excited state $|e\rangle_{\mbox{\scriptsize A}}$,
so that we can examine whether the NS gate works or fails.
The output state of the NS gate is sent to the path $a1$ of the beam splitter BS1
in figure~\ref{Figure03}.}
\label{Figure02}
\end{figure}

\begin{figure}
\centerline{\includegraphics[scale=1.0]{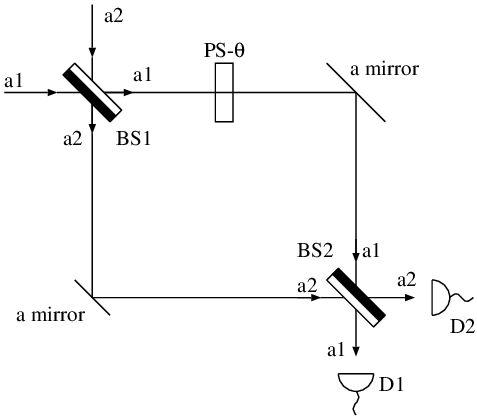}}
\caption{The Mach-Zehnder interferometer that examines the output state of the NS gate
outlined in figure~\ref{Figure02}.
Symbols BS1 and BS2 represent beam splitters (half-silvered mirrors).
A symbol PS-$\theta$ represents a phase shifter which causes the operation,
$|n\rangle_{\mbox{\scriptsize P}}\to e^{in\theta}|n\rangle_{\mbox{\scriptsize P}}$,
where $|n\rangle_{\mbox{\scriptsize P}}$ is the number state of the photons.
Symbols D1 and D2 represent detectors which count the number of incident photons.
The output state of the NS gate in figure~\ref{Figure02} is injected to the beam splitter BS1
through the path $a1$.
At the same time, we inject another coherent light into the path $a2$ of the beam splitter BS1
to let it interfere with the output state of the NS gate in figure~\ref{Figure02}.}
\label{Figure03}
\end{figure}

In this section, we propose an experimental setup in which we make use of coherent lights
to examine whether or not our scheme works.
An outline of the experimental setup is shown in figures~\ref{Figure02} and \ref{Figure03}.
(The experimental setup shown in this section aims
at examining whether or not
the proposed scheme for the NS gate really works.
The purpose of this experiment is confirming the function of the NS gate
against a certain input state,
which is given as a weak coherent state.
Thus,
this experiment does not intend to demonstrate the performance
of the proposed NS gate completely.
A complete experimental setup for the proposed NS gate is
discussed in section~\ref{an-experimental-setup2}.)

The beam splitters (half-silvered mirrors) BS1 and BS2
in figure~\ref{Figure03} transform the incident number states of paths
$a1$ and $a2$ as equation~(\ref{function-beamsplitter}).
A symbol PS-$\theta$ represents a phase shifter which causes the operation,
$|n\rangle_{\mbox{\scriptsize P}}\to e^{in\theta}|n\rangle_{\mbox{\scriptsize P}}$,
where $|n\rangle_{\mbox{\scriptsize P}}$ is the number state of the photons.
Symbols D1 and D2 represent detectors which count the number of incident photons,
so that D1 and D2 identify differences in the number states of the photons,
$|0\rangle_{\mbox{\scriptsize P}}$, $|1\rangle_{\mbox{\scriptsize P}}$,
$|2\rangle_{\mbox{\scriptsize P}}$, ... .

The experimental setup shown in figure~\ref{Figure02} works as follows.
First, we feed a strong laser beam (a coherent light) to the cavity from its left side.
Although the transmittance of the mirror of the cavity is quite low,
a few photons pass the mirror and they are reflected by the mirrors of both sides
of the cavity many times.
After this process, the photons in the cavity develop into a coherent light
of the cavity mode,
which is given by
\begin{eqnarray}
|\alpha\rangle
&=&
\exp(-|\alpha|^{2}/2)\sum_{n=0}^{\infty}\frac{\alpha^{n}}{\sqrt{n!}}
|n\rangle_{\mbox{\scriptsize P}} \nonumber \\
&=&
\exp(-|\alpha|^{2}/2)\sum_{n=0}^{\infty}
\frac{(\alpha a^{\dagger})^{n}}{n!}|0\rangle_{\mbox{\scriptsize P}} \nonumber \\
&=&
\exp(-|\alpha|^{2}/2)e^{\alpha a^{\dagger}}|0\rangle_{\mbox{\scriptsize P}},
\label{definition-coherent-light-1}
\end{eqnarray}
where $\alpha$ is an arbitrary complex number.
(Here, we provide that $0!=1$.
In the notation of the coherent states of photons,
we omit the index P.)
We can assume that this coherent light is weak, so that $|\alpha|<1$.
We let the width of the cavity (the length between the mirrors of the cavity) be equal
to a half of the wavelength of the coherent light.
Thus, the cavity mode forms a standing wave.

Second, we put the two-level atom at an anti-node of the standing wave of the cavity mode.
For example, we can capture and locate an ionized atom in a certain region by a technique
of the Paul trap (a quadrupole ion trap).\cite{Raizen1992} \ 
We can also locate the atom in a certain area by injecting a slow atomic beam there.
Then, the coherent light $|\alpha\rangle$ interacts with the atom as the JCM.
If the time of flight of the atom is equal to
$T=3\pi/(\sqrt{2}|\kappa|)$ or $T=7\pi/(\sqrt{2}|\kappa|)$,
and if we observe $|g\rangle_{\mbox{\scriptsize A}}$ with the selective electric field detector
in figure~\ref{Figure02},
an approximate NS gate is realized.
The coherent state $|\alpha\rangle$ is transformed and the reduction of the state vector is occurred
as follows:
\begin{eqnarray}
&&
|\alpha\rangle
=
\exp(-|\alpha|^{2}/2)
\Bigl[
|0\rangle_{\mbox{\scriptsize P}}+\alpha|1\rangle_{\mbox{\scriptsize P}}
+\frac{\alpha^{2}}{\sqrt{2}}|2\rangle_{\mbox{\scriptsize P}}
+O(|\alpha|^{3})|\varphi^{(1)}\rangle_{\mbox{\scriptsize P}}
\Bigr] \nonumber \\
&\longrightarrow&
|\Psi\rangle_{\mbox{\scriptsize P}}
\simeq
\exp(-|\alpha|^{2}/2)
\Bigl[
|0\rangle_{\mbox{\scriptsize P}}+|d(m)|\alpha|1\rangle_{\mbox{\scriptsize P}}
-\frac{\alpha^{2}}{\sqrt{2}}|2\rangle_{\mbox{\scriptsize P}} \nonumber \\
&&\quad\quad\quad\quad
+O(|\alpha|^{3})|\varphi^{(1)}\rangle_{\mbox{\scriptsize P}}
\Bigr],
\label{NS-gate-operation-coherent-light-1}
\end{eqnarray}
where $|\varphi^{(1)}\rangle_{\mbox{\scriptsize P}}$
is a normalized superposition of
$\{|3\rangle_{\mbox{\scriptsize P}},|4\rangle_{\mbox{\scriptsize P}},
|5\rangle_{\mbox{\scriptsize P}}, ...\}$
and
$m=1$ or $3$.
Because $d(m)\simeq\pm 1$
as shown in table~\ref{Table1},
equation~(\ref{NS-gate-operation-coherent-light-1}) holds approximately.

Here, we define the following superposition of two coherent states:
\begin{eqnarray}
&&
|\sqrt{3}e^{i\theta_{0}}\alpha\rangle
+
|\sqrt{3}e^{-i\theta_{0}}\alpha\rangle \nonumber \\
&=&
2\exp(-\frac{3}{2}|\alpha|^{2})
\Bigl[
|0\rangle_{\mbox{\scriptsize P}}+\alpha|1\rangle_{\mbox{\scriptsize P}}
-\frac{\alpha^{2}}{\sqrt{2}}|2\rangle_{\mbox{\scriptsize P}}
+O(|\alpha|^{3})|\varphi^{(2)}\rangle_{\mbox{\scriptsize P}}
\Bigr],
\label{NS-gate-operation-coherent-light-2}
\end{eqnarray}
where $\sin\theta_{0}=\sqrt{2/3}$
and
$\cos\theta_{0}=\sqrt{1/3}$,
and $|\varphi^{(2)}\rangle_{\mbox{\scriptsize P}}$ is a normalized superposition of
$\{|3\rangle_{\mbox{\scriptsize P}},|4\rangle_{\mbox{\scriptsize P}},
|5\rangle_{\mbox{\scriptsize P}}, ...\}$.
Looking at equations~(\ref{NS-gate-operation-coherent-light-1})
and (\ref{NS-gate-operation-coherent-light-2}),
we obtain
\begin{equation}
|\Psi\rangle_{\mbox{\scriptsize P}}
=
\frac{1}{2}e^{|\alpha|^{2}}
(|\sqrt{3}e^{i\theta_{0}}\alpha\rangle
+|\sqrt{3}e^{-i\theta_{0}}\alpha\rangle)
+O(|\alpha|^{3})|\varphi^{(3)}\rangle_{\mbox{\scriptsize P}},
\label{output-error-probability}
\end{equation}
where
$|\varphi^{(3)}\rangle_{\mbox{\scriptsize P}}$ is a normalized superposition of
$\{|3\rangle_{\mbox{\scriptsize P}},|4\rangle_{\mbox{\scriptsize P}},
|5\rangle_{\mbox{\scriptsize P}}, ...\}$.
Thus, we can regard the output state of the cavity shown in figure~\ref{Figure02} as
$(1/2)e^{|\alpha|^{2}}(|\sqrt{3}e^{i\theta_{0}}\alpha\rangle
+|\sqrt{3}e^{-i\theta_{0}}\alpha\rangle)$
with the error probability $O(|\alpha|^{3})$.

Finally, we have to extract the time-evolved state of the cavity field from the cavity.
Although the transmittance of the mirror of the cavity is quite low,
the time-evolved photons can go outside the cavity gradually over a long period of time.
[Strictly speaking, the time-evolved state of the cavity field can fly away from the cavity
to its either side (the right side or the left side).
However, to let the discussion be simple,
we assume that the time-evolved light goes outside of the cavity from its right side
in figure~\ref{Figure02}.]

Next, we examine the output state of the NS gate outlined in figure~\ref{Figure02}
with the Mach-Zehnder interferometer shown in figure~\ref{Figure03}.
The interferometer of figure~\ref{Figure03} works as follows.
First, we inject the output state generated by the NS gate in figure~\ref{Figure02}
into the path $a1$ of the beam splitter BS1.
At the same time, we inject another coherent light $|\alpha\rangle_{a2}$
into the path $a2$ of BS1 to let it interfere
with the output state of the NS gate in figure~\ref{Figure02}.
Second, we apply the phase shifter PS-$\theta$ to the state of the path $a1$.
Third, we apply the beam splitter BS2 to the paths $a1$ and $a2$.
Finally, we count the number of photons on each path, $a1$ and $a2$,
by the detectors D1 and D2, respectively.

Here, we follow the transformations applied to states of incident photons
in this interferometer.
Before discussing the function of the interferometer in figure~\ref{Figure03} in detail,
we consider two coherent lights $|\alpha\rangle$ and $|\beta\rangle$ injected
into the paths $a1$ and $a2$ of the beam splitter BS1, respectively.
Remembering the operation of BS1 in equation~(\ref{function-beamsplitter}) and
the definition of the coherent light in equation~(\ref{definition-coherent-light-1}),
we can describe the transformation caused by the beam splitter BS1 as follows:
\begin{eqnarray}
&&
|\alpha\rangle_{a1}|\beta\rangle_{a2}
=
\exp[-(|\alpha|^{2}+|\beta|^{2})/2]
\exp(\alpha a_{1}^{\dagger})\exp(\beta a_{2}^{\dagger})
|0\rangle_{a1}|0\rangle_{a2} \nonumber \\
&\stackrel{\mbox{\scriptsize BS1}}{\longrightarrow}&
\exp[-(|\alpha|^{2}+|\beta|^{2})/2]
\exp[\frac{\alpha}{\sqrt{2}}(a_{1}^{\dagger}+a_{2}^{\dagger})]
\exp[\frac{\beta}{\sqrt{2}}(a_{1}^{\dagger}-a_{2}^{\dagger})]
|0\rangle_{a1}|0\rangle_{a2} \nonumber \\
&&
=
\exp[-(|\alpha|^{2}+|\beta|^{2})/2]
\exp[\frac{1}{\sqrt{2}}(\alpha+\beta)a_{1}^{\dagger}]
|0\rangle_{a1} \nonumber \\
&&\quad\quad
\otimes
\exp[\frac{1}{\sqrt{2}}(\alpha-\beta)a_{2}^{\dagger}]
|0\rangle_{a2} \nonumber \\
&&
=
|\frac{1}{\sqrt{2}}(\alpha+\beta)\rangle_{a1}
|\frac{1}{\sqrt{2}}(\alpha-\beta)\rangle_{a2}.
\end{eqnarray}
Moreover, we think about how the phase shifter PS-$\theta$ works against the coherent light,
\begin{eqnarray}
|\alpha\rangle \nonumber
&\stackrel{\mbox{\scriptsize PS-}\theta}{\longrightarrow}&
\exp(-|\alpha|^{2}/2)
\sum_{n=0}^{\infty}\frac{(\alpha e^{i\theta})^{n}}{\sqrt{n!}}|n\rangle_{\mbox{\scriptsize P}} \nonumber \\
&&
=
|e^{i\theta}\alpha\rangle.
\end{eqnarray}

From the above formulas, we can describe the transformation applied to coherent states
in the interferometer as follows:
\begin{eqnarray}
&&
|\alpha\rangle_{a1}|\beta\rangle_{a2} \nonumber \\
&\stackrel{\mbox{\scriptsize BS1}}{\longrightarrow}&
|\frac{1}{\sqrt{2}}(\alpha+\beta)\rangle_{a1}
|\frac{1}{\sqrt{2}}(\alpha-\beta)\rangle_{a2} \nonumber \\
&\stackrel{\mbox{\scriptsize PS-}\theta}{\longrightarrow}&
|\frac{e^{i\theta}}{\sqrt{2}}(\alpha+\beta)\rangle_{a1}
|\frac{1}{\sqrt{2}}(\alpha-\beta)\rangle_{a2} \nonumber \\
&\stackrel{\mbox{\scriptsize BS2}}{\longrightarrow}&
|\frac{1}{2}[(e^{i\theta}+1)\alpha+(e^{i\theta}-1)\beta]\rangle_{a1}
\otimes
|\frac{1}{2}[(e^{i\theta}-1)\alpha+(e^{i\theta}+1)\beta]\rangle_{a2}.
\end{eqnarray}
Hence, the Mach-Zehnder interferometer in figure~\ref{Figure03}
performs the transformation against the output state of the cavity
$|\Psi\rangle_{\mbox{\scriptsize P}}$
given by equations~(\ref{NS-gate-operation-coherent-light-1}),
(\ref{NS-gate-operation-coherent-light-2})
and (\ref{output-error-probability})
as follows:
\begin{eqnarray}
&&
|\Psi\rangle_{\mbox{\scriptsize P}}
|\alpha\rangle_{a2}
=
\frac{1}{2}e^{|\alpha|^{2}}
(|\sqrt{3}e^{i\theta_{0}}\alpha\rangle_{a1}
+|\sqrt{3}e^{-i\theta_{0}}\alpha\rangle_{a1})
|\alpha\rangle_{a2} \nonumber \\
&&\quad\quad\quad\quad\quad\quad
+
O(|\alpha|^{3})|\varphi^{(3)}\rangle_{a1}|\alpha\rangle_{a2} \nonumber \\
&\longrightarrow&
\frac{1}{2}e^{|\alpha|^{2}}
|\frac{1}{2}\alpha F_{1}(\theta)\rangle_{a1}
|\frac{1}{2}\alpha F_{2}(\theta)\rangle_{a2}
+
\frac{1}{2}e^{|\alpha|^{2}}
|\frac{1}{2}\alpha F_{3}(\theta)\rangle_{a1}
|\frac{1}{2}\alpha F_{4}(\theta)\rangle_{a2} \nonumber \\
&&\quad\quad
+O(|\alpha|^{3})|\varphi'\rangle_{a1,a2},
\label{Mach-Zehnder-transformation-1}
\end{eqnarray}
where
\begin{eqnarray}
F_{1}(\theta)
&=&
(e^{i\theta}+1)\sqrt{3}e^{i\theta_{0}}+(e^{i\theta}-1), \nonumber \\
F_{2}(\theta)
&=&
(e^{i\theta}-1)\sqrt{3}e^{i\theta_{0}}+(e^{i\theta}+1), \nonumber \\
F_{3}(\theta)
&=&
(e^{i\theta}+1)\sqrt{3}e^{-i\theta_{0}}+(e^{i\theta}-1), \nonumber \\
F_{4}(\theta)
&=&
(e^{i\theta}-1)\sqrt{3}e^{-i\theta_{0}}+(e^{i\theta}+1),
\end{eqnarray}
and $|\varphi'\rangle_{a1,a2}$ is a certain normalized state of the paths $a1$ and $a2$.
In equation~(\ref{Mach-Zehnder-transformation-1}), the term of
$O(|\alpha|^{3})$
implies the error probability.

\begin{figure}
\centerline{\includegraphics[scale=1.0]{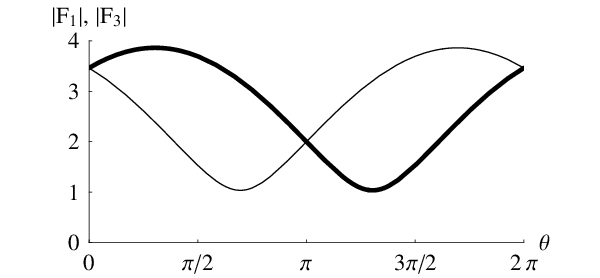}}
\caption{Graphs of $|F_{1}(\theta)|$ and $|F_{3}(\theta)|$ for $\theta\in[0,2\pi]$.
A thick curve represents $|F_{1}(\theta)|$
and a thin curve represents $|F_{3}(\theta)|$.
Looking at these graphs, we notice the following facts.
When $\theta=\arccos(-1/3)\simeq 1.911$
and
$\theta=2\pi-\arccos(-1/3)\simeq 4.372$,
the difference between $|F_{1}(\theta)|$ and $|F_{3}(\theta)|$
reaches the maximum value.}
\label{Figure04}
\vfill
\centerline{\includegraphics[scale=1.0]{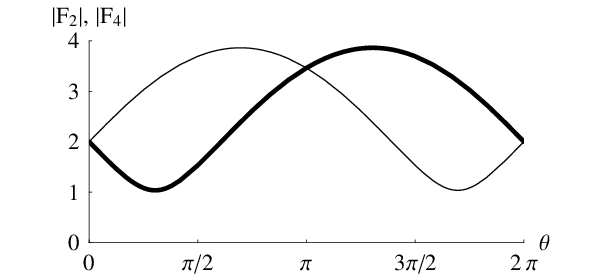}}
\caption{Graphs of $|F_{2}(\theta)|$ and $|F_{4}(\theta)|$ for $\theta\in[0,2\pi]$.
A thick curve represents $|F_{2}(\theta)|$
and a thin curve represents $|F_{4}(\theta)|$.
Looking at these graphs, we notice the following facts.
When $\theta=\arccos(1/3)\simeq 1.231$
and
$\theta=2\pi-\arccos(1/3)\simeq 5.052$,
the difference between $|F_{2}(\theta)|$ and $|F_{4}(\theta)|$
reaches the maximum value.}
\label{Figure05}
\end{figure}

We plot $|F_{1}(\theta)|$ and $|F_{3}(\theta)|$ for $\theta\in[0,2\pi]$ in figure~\ref{Figure04},
and
$|F_{2}(\theta)|$ and $|F_{4}(\theta)|$ for $\theta\in[0,2\pi]$ in figure~\ref{Figure05}.
Looking at figures~\ref{Figure04} and \ref{Figure05},
we pay attention to the following facts.
When $\theta=\arccos(1/3)$,
the difference between
$|F_{2}(\theta)|$ and $|F_{4}(\theta)|$ reaches the maximum value.
Then,
we obtain
$|F_{1}(\arccos(1/3))|=2\sqrt{11/3}$,
$|F_{2}(\arccos(1/3))|=2/\sqrt{3}$,
$|F_{3}(\arccos(1/3))|=2$,
and $|F_{4}(\arccos(1/3))|=2\sqrt{3}$.

The right-hand side of equation~(\ref{Mach-Zehnder-transformation-1}) shows
a sign of weak entanglement between the paths $a1$ and $a2$.
(This weak entanglement is generated by the photon bunching.)
Thus, if we repeat the experiment in figures~\ref{Figure02} and \ref{Figure03}
many times and store the statistical data observed with the detectors D1 and D2,
we can obtain some information about the function of the cavity shown in figure~\ref{Figure02}.

\begin{figure}
\centerline{\includegraphics[scale=1.0]{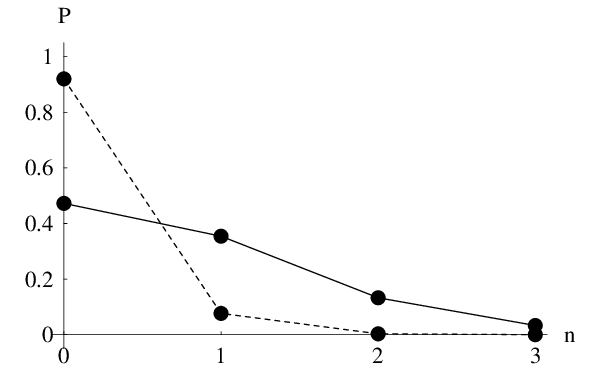}}
\caption{Graphs of the Poisson distribution functions,
$P(n,\mu)=e^{-\mu}\mu^{n}/n!$ for $n=0,1,2,3$
with
$\mu=0.75$ and $\mu\simeq 0.083\mbox{ }33$.
(The variable $\mu$ represents a mean.)
A solid line represents $P(n,0.75)$
and
a broken line represents $P(n,0.083\mbox{ }33)$.
We note that
$P(1,0.75)\simeq 0.3543$
and
$P(1,0.083\mbox{ }33)\simeq 0.076\mbox{ }67$.
We also note that
$P(2,0.75)\simeq 0.1329$
and
$P(2,0.083\mbox{ }33)\simeq 0.003\mbox{ }194$.}
\label{Figure06}
\end{figure}

Here, for example, we assume $\alpha=1/2$ and $\theta=\arccos(1/3)$.
We obtain
$|(1/2)\alpha F_{1}(\theta)|^{2}=11/12\simeq 0.9167$,
$|(1/2)\alpha F_{2}(\theta)|^{2}=1/12\simeq 0.083\mbox{ }33$,
\\
$|(1/2)\alpha F_{3}(\theta)|^{2}=1/4=0.25$,
and
$|(1/2)\alpha F_{4}(\theta)|^{2}=3/4=0.75$.
In this case, the error probability $O(|\alpha|^{3})$ is given by $0.125$ around.
If we perform photon counting against
$|(1/2)\alpha F_{4}(\theta)\rangle_{a2}$,
we obtain the Poisson distribution whose mean is given by
$|(1/2)\alpha F_{4}(\theta)|^{2}=0.75$.
If we perform photon counting against
$|(1/2)\alpha F_{2}(\theta)\rangle_{a2}$,
we obtain the Poisson distribution whose mean is given by
$|(1/2)\alpha F_{2}(\theta)|^{2}=1/12\simeq 0.083\mbox{ }33$.
We plot these distribution functions in figure~\ref{Figure06}.

If we detect one photon by the detector D2,
we become aware of reduction of the state vector on the path $a1$ to
$|(1/2)\alpha F_{3}(\theta)\rangle_{a1}$
with a comparatively high probability.
[We describe the Poisson distribution with a mean $\mu$ as
\begin{equation}
P(n,\mu)=e^{-\mu}\frac{\mu^{n}}{n!}
\quad\quad
\mbox{for $n=0,1,2,...$}.
\end{equation}
Because
$P(1,0.75)\simeq 0.3543$,
$P(1,0.083\mbox{ }33)\simeq 0.076\mbox{ }67$
and
\\
$P(1,0.75)>P(1,0.083\mbox{ }33)$,
we can expect that
the detection of one photon with D2 indicates the distribution of
$|(1/2)\alpha F_{4}(\theta)\rangle_{a2}$,
where
$|(1/2)\alpha F_{4}(\theta)|^{2}=0.75$.
Hence, we can expect that
this reduction of the state vector occurs
with comparatively high probability.
The similar things happen when we detect two photons by the detector D2.]

In this case, the probability that the detector D2 observes one photon
is given by $P(1,0.75)(e^{|\alpha|^{2}}/2)^{2}\simeq 0.1460$ around.
This probability is comparable with the error probability
$O(|\alpha|^{3})\sim 0.125$,
so that we need careful data analysis.
However,
repeating the experiment many times and storing the statistical data
which is a set of events conditioned on the detection of $|1\rangle_{a2}$
(or $|2\rangle_{a2}$)
by the detector D2,
we obtain the Poisson distribution that comes from the coherent state
$|(1/2)\alpha F_{3}(\theta)\rangle_{a1}$
with the error probability $O(|\alpha|^{3})$.
[Because $|(1/2)\alpha F_{3}(\theta)|^{2}=0.25$
is smaller than unity,
we have to store huge amounts of data of experiments for detecting
$|(1/2)\alpha F_{3}(\theta)\rangle_{a1}$.]

Hence, we can experimentally examine
whether or not the NS gate realized in figure~\ref{Figure02} works
with the interferometer shown in figure~\ref{Figure03}.

\section{\label{an-experimental-setup2}An experimental setup for constructing the NS gate
using an optical loop circuit}
In section~\ref{an-experimental-setup1},
we discuss the NS gate which only accepts the coherent state as an input.
However, to construct the genuine NS gate,
we have to put an arbitrary superposition of
$\{|0\rangle_{\mbox{\scriptsize P}},
|1\rangle_{\mbox{\scriptsize P}},
|2\rangle_{\mbox{\scriptsize P}}\}$
into the cavity,
let it interact with the two-level atom,
and extract the time-evolved state of the photons from the cavity.
In this section, we try to show another experimental setup for carrying out these procedures.
We make an optical loop circuit out of a polarizing beam splitter and the Pockels cell
in the cavity
and we let an arbitrary superposition of
$\{|0\rangle_{\mbox{\scriptsize P}},
|1\rangle_{\mbox{\scriptsize P}},
|2\rangle_{\mbox{\scriptsize P}}\}$
develop into the cavity mode.
In this method, for building the optical loop,
the polarization degree of freedom of photons plays an important role.
[Kwiat {\it et al}. construct an optical loop circuit from the polarizing beam splitters
and the Pockels cells actually in their experiment.\cite{Kwiat1999}]

First, we introduce the polarization degree of freedom to photons.
We describe photons' state vectors as follows:
$\{|n\rangle_{\mbox{\scriptsize P}}|V\rangle_{\mbox{\scriptsize P}},
|n\rangle_{\mbox{\scriptsize P}}|H\rangle_{\mbox{\scriptsize P}}:n=0,1,2,...\}$,
where $|V\rangle_{\mbox{\scriptsize P}}$ and $|H\rangle_{\mbox{\scriptsize P}}$
imply the photons' vertical and horizontal
polarization states, respectively.

Second, we prepare two kinds of devices which apply unitary transformations
to photons' polarization states:
a polarizing beam splitter and the Pockels cell.
(We let PBS and PC be the symbols of the polarizing beam splitter
and the Pockels cell, respectively.)

\begin{figure}
\centerline{\includegraphics[scale=1.0]{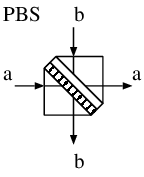}}
\caption{The polarizing beam splitter (PBS).
It has two paths $a$ and $b$.}
\label{Figure07}
\end{figure}

We draw the PBS in figure~\ref{Figure07}.
The PBS splits an unpolarized light into beams of differing polarization
(two orthogonal linearly polarized states, that is,
vertical and horizontal polarization states).
For example, we can utilize the Wollaston prism as the PBS.
Writing the PBS's incoming
and outgoing paths $a$ and $b$ as states,
\begin{equation}
|a\rangle
=
\left(
\begin{array}{c}
1 \\
0
\end{array}
\right),
\quad\quad
|b\rangle
=
\left(
\begin{array}{c}
0 \\
1
\end{array}
\right),
\label{PBS-paths-states}
\end{equation}
we can describe a unitary transformation that the PBS applies to polarized photons as
\begin{equation}
U_{\mbox{\scriptsize PBS}}
=
\left(
\begin{array}{cc}
|V\rangle\langle V| & |H\rangle\langle H| \\
|H\rangle\langle H| & |V\rangle\langle V|
\end{array}
\right).
\label{PBS-unitary-matrix}
\end{equation}
(From now on, to let the notation be simple,
we omit the index P from $|V\rangle_{\mbox{\scriptsize P}}$ and $|H\rangle_{\mbox{\scriptsize P}}$.)
For instance, if we inject $(c_{V}|V\rangle+c_{H}|H\rangle)$ into the PBS
from the path $a$,
$c_{V}|V\rangle$ and $c_{H}|H\rangle$ are separated and run away from the PBS through the paths
$a$ and $b$, respectively.

\begin{figure}
\centerline{\includegraphics[scale=1.0]{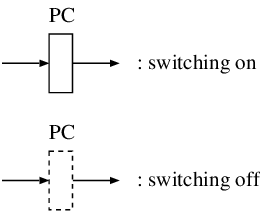}}
\caption{The Pockels cell (PC).
It has only one path.}
\label{Figure08}
\end{figure}

We draw the PC in figure~\ref{Figure08}.
The PC is a voltage-controlled wave plate.
If we apply the voltage to the PC (switching on),
polarized photons' states injected into the PC
are transformed as $|V\rangle\to|H\rangle$ and $|H\rangle\to|V\rangle$,
and they run away from the PC.
If we do not apply the voltage to the PC (switching off),
the PC leaves injected polarized photons' states untouched and returns them as outputs.

\begin{figure}
\centerline{\includegraphics[scale=1.0]{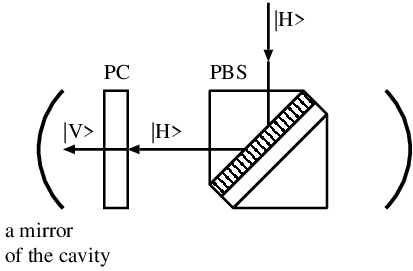}}
\caption{An optical circuit built in the cavity.
In this case, the PC is switched on.
The initially polarized photons $|H\rangle$ is put into PBS from its upper side,
and the PBS reflects $|H\rangle$ to its left side.
The PC transforms the incoming $|H\rangle$ into the outgoing $|V\rangle$.
Then, the mirror on the left side of the cavity reflects $|V\rangle$.}
\label{Figure09}
\end{figure}

\begin{figure}
\centerline{\includegraphics[scale=1.0]{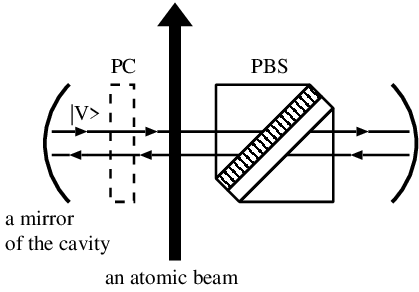}}
\caption{An optical circuit built in the cavity.
In this case, the PC is switched off.
Because the mirrors on the both sides of the cavity
reflect $|V\rangle$,
the optical circuit forms a closed loop.
The photons polarized as $|V\rangle$ fly along the closed loop many times,
and they become a cavity field.
In the area where the atom is located, the cavity field causes
the Jaynes-Cummings interaction with the atom.}
\label{Figure10}
\end{figure}

\begin{figure}
\centerline{\includegraphics[scale=1.0]{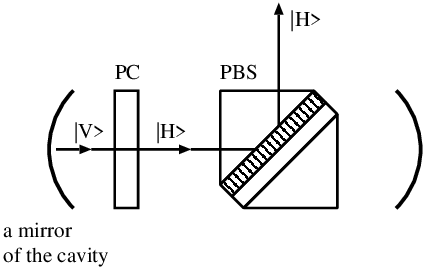}}
\caption{An optical circuit built in the cavity.
In this case, the PC is switched on.
After reflected by the mirror on the left side of the cavity,
$|V\rangle$ flies against the PC and the PC transforms $|V\rangle$ into $|H\rangle$.
And then, the polarized photons' state $|H\rangle$
is reflected by the PBS to its upper side,
so that the photons run away from the cavity.}
\label{Figure11}
\end{figure}

Third, we construct an optical circuit in the cavity
as shown in figures~\ref{Figure09}, \ref{Figure10}
and \ref{Figure11}.
We inject the photons' wave function initialized as $|H\rangle$
into the PBS and it is reflected by the PBS as shown in figure~\ref{Figure09}.
At this moment, we switch on the PC,
so that the photons' state $|H\rangle$ is transformed into $|V\rangle$.
Then, $|V\rangle$ flies against the mirror of the cavity in the left side
and it is reflected by the mirror.
Next moment, we switch off the PC as shown in figure~\ref{Figure10}.
Because the photons' state $|V\rangle$ passes across the PBS,
it is reflected by the mirrors on the both sides of the cavity
and it runs along an optical closed loop.
While the photons in the state of $|V\rangle$ run along this loop many times,
they develop into the cavity mode.
This cavity mode causes the Jaynes-Cummings interaction
with the two-level atom which is injected into the cavity as a slow atomic beam.
After a period for constructing the NS gate
[$T=3\pi/(\sqrt{2}|\kappa|)$ or $7\pi/(\sqrt{2}|\kappa|)$],
we switch on the PC as shown in figure~\ref{Figure11}.
Because the PC transforms the photons' state $|V\rangle$
reflected by the mirror on the left side of the cavity
into $|H\rangle$,
the photons are reflected by the PBS  and they run away from the cavity.

In general,
the pulsed photons injected into the NS gate are given as a wave packet.
Its shape depends on a certain dispersion relation.
Thus,
the wavelength (the frequency) of the photons is given
by a probability distribution.
When the wave packet is fed into the cavity,
the photons, which make a dominant contribution
to the probability distribution of wavelengths, gradually change
into a single cavity mode
and they interact with the atom as the JCM.
(Describing the wave packet as a sum of the Fourier components,
the photons with the mean wavelength,
which make a major contribution to the probability distribution,
evolve into a single cavity mode.
Strictly speaking,
the pulsed photons contain various components of wavelengths.
Thus,
some components that are far from the mean value cause
minor nonlinear effects to the evolution of a cavity mode.
However,
such a rigorous treatment is beyond the purpose of this paper,
so that we neglect these effects for simplicity.)

The above is the outline for constructing the cavity mode with an optical circuit.
From the theoretical viewpoint,
our optical circuit works in principle.
However, examining each optical device in the circuit,
we notice some problems for us to perform this experiment actually.
Here, we try to go into details about each device.

To construct the closed loop, we have to turn on and off the Pockels cell
a few times at short intervals.
At the present time, the Pockels cell that has a $2.5\times 10^{-10}$ s
time response has been developed.
In the experiment of the JCM with the cavity quantum electrodynamics system
carried out by Rempe {\it et al}.,\cite{Rempe1987}
$63p_{3/2}\leftrightarrow 61d_{5/2}$ transition of ${}^{85}\mbox{Rb}$
($f=21\mbox{ }456.0\times 10^{6}$ Hz,
$\lambda=1.397\mbox{ }24\times 10^{-2}$ m)
is made use of for the two-level atom,
and their cavity gives the coupling constant
$|\kappa|\simeq(1/70)\times 10^{6}$ $\mbox{s}^{-1}$.
If we construct the optical loop circuit in the cavity whose width is given by
$L=\lambda/2=6.986\times 10^{-3}$ m,
the time response required to the PC is estimated at
$L/c=2.330\times 10^{-11}$ s,
where $c=2.998\times 10^{8}$ m$\mbox{s}^{-1}$ is a velocity of the light.
Thus, at present,
we cannot prepare such an ultra-fast Pockels cell which is useful in our experiment.
If we construct the cavity whose
coupling constant is similar to that in
Rempe {\it et al}.'s experiment,
the period of the time evolution for the JCM is estimated to be
$T=3\pi/(\sqrt{2}|\kappa|)\simeq 4.67\times 10^{-4}$ s for $m=1$ and
$T=7\pi/(\sqrt{2}|\kappa|)\simeq 1.09\times 10^{-3}$ s for $m=3$.

In figure~\ref{Figure10},
the photons pass across the PC and the PBS.
At present, the rate of the insertion loss of the PC is given by $0.04$ around.
In contrast, the rate of the insertion loss of the PBS is lower than $0.01$.
Thus, in our optical loop, cavity loss due to dissipative effect in the Pockels cell is very serious.

Finally, we have to point out a fact that to make a small optical circuit in the cavity is
difficult even with the latest technology.
If we build the optical circuit in the cavity,
its size may be around a few centimeters.
(The size of available optical devices on the market is around a few centimeters.)
The coupling constant $|\kappa|$ is proportional to $\sqrt{\omega/L^{3}}$,
where $L^{3}$ is the volume of the cavity.
Hence, to make a small cavity is favourable to us.
In Rempe {\it et al}.'s experimental setup,
photons' wavelength is given by
$\lambda=1.397\mbox{ }24\times 10^{-2}$ m
($f=21\mbox{ }456.0\times 10^{6}$ Hz).\cite{Rempe1987} \ 
Thus, the length between mirrors in the cavity
has to be equal to $\lambda/2=6.986\times 10^{-3}$ m.

\section{\label{discussions}Discussions}
In this paper, we discuss how to build Knill, Laflamme and Milburn's
nonlinear sign-shift gate with the Jaynes-Cummings model.
We also discuss the experimental setups for our scheme.
The first one of our experimental setups seems to be practical and easy to carry out
in the laboratory
because it utilizes coherent lights.
In contrast, the second one of our experimental setups seems to be difficult to demonstrate
actually
because it requires optical devices that have very excellent performance.
However, because optical devices' performance is improving rapidly,
the author thinks that our second experimental setup will be demonstrated in the near future.

As we mentioned in section~\ref{introduction},
although the Jaynes-Cummings model was born about forty years ago,
it is studied from the new viewpoint by the researchers
of the quantum information science.\cite{Bose2001,Scheel2003,
Azuma2008-1,Yu2004,Yonac2006,Yu2006,Almeida2007,Gilchrist2003,Marchiolli2003,Marchiolli2006} \
The author thinks that we can find many new applications from the Jaynes-Cummings model.

\section*{Acknowledgements}
The author thanks K.~Kuwahara and colleagues of iCFD for encouragement.
The author also thanks W.~J.~Munro and M.~A.~Marchiolli for drawing his attention
to references~\citen{Gilchrist2003,Marchiolli2003,Marchiolli2006}.
The author thanks N.~Hatano and T.~Sagawa for critical reading of the manuscript.

%\appendix
%\section{First Appendix} %Empty argument \section{} yields `Appendix'. 
%
%\section{Second Appendix}

\end{document}